\documentclass[sigplan,10pt,screen,authorversion,nonacm]{acmart}
\usepackage{preamble}
\usepackage{lang-defs}

\makeatletter
\def\@ACM@checkaffil{
    \if@ACM@instpresent\else
    \ClassWarningNoLine{\@classname}{No institution present for an affiliation}%
    \fi
    \if@ACM@citypresent\else
    \ClassWarningNoLine{\@classname}{No city present for an affiliation}%
    \fi
    \if@ACM@countrypresent\else
        \ClassWarningNoLine{\@classname}{No country present for an affiliation}%
    \fi
}
\makeatother

\newcommand\blfootnote[1]{%
  \begingroup
  \renewcommand\thefootnote{}\footnote{#1}%
  \addtocounter{footnote}{-1}%
  \endgroup
}

\setitemize{left=1pt..1.5em}
\setenumerate{left=1pt..1.5em}

\definecolor{olivegreen}{RGB}{0,153,0}
\definecolor{awesome}{rgb}{1.0, 0.13, 0.32}
\mdfdefinestyle{mystyle}{
    backgroundcolor=white!20
}
\definecolor{ygreen}{HTML}{d6e39d}
\definecolor{sgreen}{HTML}{bee69c}
\definecolor{arylideyellow}{rgb}{0.91, 0.84, 0.42}
\definecolor{citrine}{rgb}{0.89, 0.82, 0.04}
\definecolor{Cerulean}{RGB}{197.2,255.0,113.5}

\newif\ifcorrectingmode
\correctingmodetrue
\LetLtxMacro\origcite\cite

\def \editMode {True}

\ifx\editMode\undefined
    \newcommand{\del}[1]{}
    \newcommand{\fixme}[1]{}
    \newcommand{\urkfixme}[1]{}
    \newcommand{\vk}[1]{}
    \newcommand{\ra}[1]{}
    \newcommand{\sj}[1]{}
    \newcommand{\albert}[1]{}
    \newcommand{\fixmecleanuppl}[2][]{}
\else
    \newcommand{\del}[1]{}
    \newcommand{\fixme}[1]{\textcolor{red}{\textbf{URK:$\bigstar$}#1}}
    \newcommand{\urkfixme}[1]{\textcolor{red}{\textbf{URK FIXME}#1}}
    \newcommand{\vk}[1]{\textcolor{blue}{\textbf{VK:$\bigstar$}#1}}
    \newcommand{\sj}[1]{\textcolor{orange}{\textbf{SJ:$\bigstar$}#1}}
    \newcommand{\albert}[1]{\textcolor{brown}{\textbf{Albert:$\bigstar$}#1}}

    \newcommand{\fixmecleanuppl}[2][]{\fixme{NEEDS WORK}}
\fi

\newcommand{\rlforreal}[1][]{{\sc RL4ReAl}\xspace}
\newcommand{\posetrl}[1][]{{POSET-RL}\xspace}
\newcommand{\posetrlpass}[1][]{\texttt{PosetRL}\xspace}
\newcommand{\loopdist}[1][]{{RL-LoopDistribution}\xspace}
\newcommand{\inliner}[1][]{{Inliner}\xspace}

\newcommand{\specI}[1][]{{SPEC CPU 2006}\xspace}
\newcommand{\specII}[1][]{{SPEC CPU 2017}\xspace}
\newcommand{\polybench}[1][]{{PolyBench}\xspace}

\newcommand{\grpc}{gRPC\xspace}
\newcommand{\onnx}{ONNX\xspace}
\newcommand{\pipe}{Pipe\xspace}
\newcommand{\tensorflow}{TensorFlow\xspace}
\newcommand{\tf}{\tensorflow}

\newcommand{\protobuf}{Protobuf\xspace}
\newcommand{\json}{JSON\xspace}
\newcommand{\bitstream}{Bitstream\xspace}

\newcommand{\mlRunner}{\texttt{MLModelRunner}\xspace}
\newcommand{\mlRunners}{MLModelRunners\xspace}
\newcommand{\grpcRunner}{\texttt{gRPCModelRunner}\xspace}
\newcommand{\pipeRunner}{\texttt{pipeModelRunner}\xspace}
\newcommand{\onnxRunner}{\texttt{ONNXModelRunner}\xspace}

\newcommand{\serdes}{\texttt{SerDes}\xspace}

\newcommand{\serdesntt}{SerDes\xspace}
\newcommand{\baseSerDes}{\texttt{BaseSerDes}\xspace}
\newcommand{\protoSerDes}{\texttt{ProtobufSerDes}\xspace}
\newcommand{\jsonSerDes}{\texttt{JSONSerDes}\xspace}
\newcommand{\bitSerDes}{\texttt{BitstreamSerDes}\xspace}

\newcommand{\mlmodel}{ML Model\xspace}

\newcommand{\toolname}{\textsc{ML-Compiler-Bridge}\xspace}
\definecolor{lgreen}{HTML}{E6FFCC}

\AtBeginDocument{%
  \providecommand\BibTeX{{%
    \normalfont B\kern-0.5em{\scshape i\kern-0.25em b}\kern-0.8em\TeX}}}

\acmYear{2024}
\setcopyright{rightsretained}
\acmConference[]{}{}{}

\begin{document}
\title{The Next 700 ML-Enabled Compiler Optimizations}

\author{S. VenkataKeerthy}
\orcid{0000-0003-1393-7321}
\affiliation{
  \institution{IIT Hyderabad, India}
}

\author{Siddharth Jain}
\orcid{0000-0003-3801-7759}
\affiliation{
  \institution{IIT Hyderabad, India}
}

\author{Umesh Kalvakuntla}
\orcid{0009-0008-1807-6712}
\affiliation{
  \institution{IIT Hyderabad, India}
}

\author{Pranav Sai Gorantla}
\orcid{0009-0003-1194-2546}
\affiliation{
  \institution{IIT Hyderabad, India}
}

\author{Rajiv Shailesh Chitale}
\orcid{0009-0000-1745-6328}
\affiliation{
  \institution{IIT Hyderabad, India}
}

\author{Eugene Brevdo}
\orcid{0009-0005-7965-3534}
\affiliation{%
  \institution{Google DeepMind, USA}
}

\author{Albert Cohen}
\orcid{0000-0002-8866-5343}
\affiliation{%
  \institution{Google DeepMind, France}
}

\author{Mircea Trofin}
\orcid{0000-0002-4716-3400}
\affiliation{%
  \institution{Google, USA}
 }

\author{Ramakrishna Upadrasta}
\orcid{0000-0002-5290-3266}
\affiliation{%
  \institution{IIT Hyderabad, India}
}

\renewcommand{\shortauthors}{S. VenkataKeerthy et al.}

\begin{abstract}
    There is a growing interest in enhancing compiler optimizations with ML models, yet interactions between compilers and ML frameworks remain challenging.
Some optimizations require tightly coupled models and compiler internals, raising issues with modularity, performance and framework independence.
Practical deployment and transparency for the end-user are also important concerns.
We propose \toolname{} to enable ML model development within a traditional Python framework while making end-to-end integration with an optimizing compiler possible and efficient. We evaluate it on both research and production use cases, for training and inference, over several optimization problems, multiple compilers and its versions, and gym infrastructures. 

\end{abstract}

\maketitle
\blfootnote{Intial version prepared on 1st Sep 2023. Revised on 14th Nov 2023.}

\section{Introduction}
\label{sec:introduction}

With the success of Machine Learning (ML) models in various domains, there is a growing interest in applying ML to improve optimization heuristics in compilers~\cite{Cooper-Adaptive21stCentury, Compilers2.0Amarasinghe20}.
Several ML and Reinforcement Learning (RL) approaches have been proposed to improve optimizations like vectorization~\cite{hajali2020neurovectorizer, mendis2019ImitationLearning}, loop unrolling, distribution~\cite{Stephenson2005, shalini-rl-loop-distribution-2022}, 
function inlining~\cite{Simon-Inlining-10.1109/CGO.2013.6495004, trofin20MLGO}, register allocation~\cite{das2020, mlgo-regalloc-rfc, kim2022pbqp, VenkataKeerthy-2023-RL4ReAl}, 
prediction of phase sequences~\cite{MiCOMP, AutoPhaseHuang2019, Jain-POSET-RL-2022}, among many others~\cite{allamanis2018survey,wangSurvey2018}.
More specifically, the widely used LLVM compiler~\cite{Lattner:2004:llvm} has support for RL-based inlining decisions from version 11, and RL-based eviction decisions in its register allocator from version 14~\cite{mlgo-regalloc-rfc}.
The title of our paper acknowledges this growing trend and anticipates the needs of the ML-enabled optimizations that are yet to come, in the spirit of Landis' seminal paper~\cite{10.1145/365230.365257} on the diversity of existing and future programming languages.

Setting up an ML-based compiler optimization is a challenging task. In addition to model design, it involves specialized data collection, compiler engineering, packaging:

\begin{enumerate}
    \item Preparing or generating the data sets for training the model~\cite{cummins2017synthBenchmark, anghabench2021}.
    
    \item Engineering objective-specific features~\cite{Fursin2011}, or extracting objective-independent program embeddings~\cite{alon2019code2vec, ncc, VenkataKeerthy-2020-IR2Vec, cummins2021PrograMl}, or a combination of both.

    \item Setting up a training interface with the compiler, with examples ranging from communicating the output via compiler flags \cite{Jain-POSET-RL-2022}, offline file logs \cite{hajali2020neurovectorizer}, to generic gym APIs~\cite{OpenAIGym2016} and recent compiler-specific gym APIs like CompilerGym~\cite{cummins2021compilergym}, PolyGym~\cite{polygym2021pact} and Supersonic~\cite{supersonicRLgym2022cc}. 

    \item Finally, building and deploying the compiler with the trained model for inference.
\end{enumerate}

In most works, the process ends with step (3) and a simplified benchmark-oriented version of step (4) to evaluate the trained model. 
Indeed, while there exist a number of solutions for steps (1 \& 2), 
a proper methodology based solutions for steps (3) \& (4) that involve model-compiler interaction between have not yet been adequately addressed.

\begin{table*}[h!tb]
\centering
\caption{Diverse ML and RL requirements in previous work; unknown or unclear ones are left blank.}
\label{tab:rel-works-comp}
\resizebox{\textwidth}{!}{%
\begin{tabular}{llllllll}
\toprule
 & \multicolumn{1}{c}{\textbf{Communication}} & \multicolumn{1}{c}{\textbf{Model Input}} & \multicolumn{1}{c}{\textbf{Model Output}} & \multicolumn{1}{c}{\textbf{Commn Freq}} & \multicolumn{1}{c}{\textbf{\#Agents}} & \multicolumn{1}{c}{\textbf{Model Type}} & \multicolumn{1}{c}{\textbf{ML Framework}} \\
 \midrule
\textbf{SLP Vectorization~\cite{mendis2019ImitationLearning}} &  & LLVM IR & Instructions to pack &  & Single Agent & GGNN &  \\

\textbf{NeuroVectorizer~\cite{hajali2020neurovectorizer}} & Pragmas in source & Code2Vec vectors & \begin{tabular}[c]{@{}l@{}}source code with pragma \\ added in Python\end{tabular} & Once, at the end & Single Agent & FCNN & Keras, RLLib \\
\textbf{Register Allocation~\cite{das2020}} & No integration & Interference graph & Coloured IG & None & NA & LSTM & TensorFlow \\
\textbf{Register Allocation~\cite{kim2022pbqp}} & & PBQP Graph & Allocated PBQP graph & & Single Agent & GCN, ResNet & Pytorch \\
\textbf{POSET-RL~\cite{Jain-POSET-RL-2022}} & Opt Flags & IR2Vec vectors & Pass sequence & \begin{tabular}[c]{@{}l@{}}Multiple times \\ per episode\end{tabular} & Single Agent & FCNN & PyTorch \\
\textbf{Loop Distribution~\cite{shalini-rl-loop-distribution-2022}} & Python Wrappers & IR2Vec vectors & Distribution sequence & Once, at the end & Two agents & GNN, FCNN & PyTorch \\
\textbf{Inliner~\cite{trofin20MLGO}} & Precompiled TF model & Features & Yes/No & Once, at the end & Single Agent & FCNN & TensorFlow \\
\textbf{RegAlloc Eviction~\cite{mlgo-regalloc-rfc}} & Precompiled TF model & Features & Index of Live Range to Evict & Once, at the end & Single Agent & FCNN & TensorFlow \\
\textbf{RL4ReAl~\cite{VenkataKeerthy-2023-RL4ReAl}} & gRPC & \begin{tabular}[c]{@{}l@{}}IG + node level \\ embeddings\end{tabular} & Color map & \begin{tabular}[c]{@{}l@{}}Multiple times \\ per episode\end{tabular} & \begin{tabular}[c]{@{}l@{}}Four agents; \\ hierarchical\end{tabular} & GNN, FCNN & PyTorch, RLLib \\
\bottomrule
\end{tabular}%
}
\vskip-.5\baselineskip
\end{table*}

The diversity of compiler optimizations and ML models is associated with an equally broad range of requirements for model-compiler interaction. In Tab.~\ref{tab:rel-works-comp}, we illustrate this on recent proposals.
There exists multiple ML frameworks and even more types of ML models.
A model's input may be a plain floating point vector, or tensors of different ranks and shapes.
Outputs range from a unique Boolean decision to complex data structures.
These need to be communicated with the compiler; it may be only once for simple scenarios, or many times and involving large amounts of data for more intricate ones.
And this may involve extensive source code modifications for the sole purpose of implementing the compiler-model interface.

Some of these interactions have been explored in the literature and even landed in production; however, there \emph{does not exist} a single generic method to address the vast diversity of scenarios that are imaginable and the trade-offs therein.
Such a situation limits the scope, applicability and effectiveness of ML for compiler optimizations in the following ways:
\begin{itemize}
    \item \textbf{Scalability:} Integrating a Python model with C++ code using wrappers induces significant~\cite{shalini-rl-loop-distribution-2022} compile time overhead: e.g.\ $6\times$--$100\times$.
    
    \item \textbf{Integration:} Not all optimizations are simple enough that the outputs of the model can be communicated using flags~\cite{VenkataKeerthy-2023-RL4ReAl, shalini-rl-loop-distribution-2022, kim2022pbqp, trofin20MLGO}.
    As ML-based optimizations grow in popularity, flag-based approaches become unwieldy.

    \item \textbf{Programmability:} ML models are typically written in Python across different frameworks like TensorFlow, JAX, PyTorch, etc. 
    Expecting the model to be written \emph{in C++} \emph{within} the compiler is \emph{not ML developer-friendly}.
    
    \item  \textbf{Portability:} Several proposals involve a \emph{tight coupling} between the compiler and a specific ML framework; we however believe that a generic compiler infrastructure like LLVM \textit{should remain ML-framework-independent}.
\end{itemize}

The existing gym libraries primarily aim at facilitating model training for research and reproducibility by providing a high-level integration.  
For example, the recent CompilerGym~\cite{cummins2021compilergym} provides a high-level interface in the form of C++ wrapper methods \textit{outside} the compiler to invoke \textit{out-of-tree compiler APIs} to materialize the predicted actions. 
Such integration caters well to training certain interactions like Phase Ordering~\cite{Jain-POSET-RL-2022}. 
However, other optimizations like RegAlloc~\cite{VenkataKeerthy-2023-RL4ReAl, kim2022pbqp, das2020}, loop distribution~\cite{shalini-rl-loop-distribution-2022} and inlining~\cite{trofin20MLGO} necessitate a \textit{deeper} interfacing of the model within the compiler; with multiple rounds of interaction for both training and inference scenarios.
Further, in these gym libraries, the inference flow is driven by Python: the compilation starts by invoking a Python process, breaking the isolation between the end user and the internal compiler algorithms; this limits deployment opportunities among other downsides. We discuss these issues in detail in Sec.~\ref{sec:ml-llvm-project}.

To address these shortcomings, we propose \toolname{}, a library that allows ML model development within a traditional Python framework while providing tightly coupled and efficient end-to-end integration with the compiler.
Our library bridges the compiler and ML model by providing a suite of communication approaches (model runners) and the related (de-)serialization mechanisms (\serdesntt) to cater to diverse scenarios. 
It also provides support for both inter- and in-process communication by exposing different model runners: \grpc and named-pipes for the former, and the \tf interface and \onnx for the latter.
Diverse \serdesntt options based on \protobuf, \json, and native bitstreams improve efficiency and versatility.
The appropriate model runner and \serdesntt can be chosen based on the usage scenario and requirement, and these may differ during training and inference. Our library provides C++ and Python APIs to expose model runners and \serdesntt for integration with compilers and ML frameworks respectively. 

We show that the inter-process model runners effectively supports training. Once the model is trained, the in-process model runners provide interfacing of the model within the compiler in a transparent manner, with much lesser latency to aid in deployment.
Besides, our both model runner and \serdesntt modules can be easily extended to support more forms of communication and serialization.
Our library also provides C-APIs to aid in integration with C-based compiler infrastructures like Pluto, GCC, and SQLite.

We evaluate \toolname{} on four ML-enabled optimizations in LLVM: \loopdist, \posetrl, \rlforreal, and Inliner. We show that our library can be integrated with other compilers like Pluto~\cite{bondhugula-pluto-2008} and MLIR~\cite{lattner-mlir-2021} with minimal effort. We study the impact of communication and serialization options on compile time under different complex scenarios that the existing infrastructures could not handle.
We conduct extensive evaluations to measure the overhead caused by each model runner and \serdesntt. We also study the impact of integrating \toolname{} with LLVM in terms of additional dependencies, compile-time, and binary size overhead. Here are our contributions:
\begin{itemize}
    \item We propose \toolname{}, a library to enable the deeper integration of ML models and the compiler in a framework-independent manner.
    \item We provide a suite of \textit{two}-inter- and \textit{two}-in-process model runners, and \textit{three} (de-)serialization mechanisms (\serdesntt) to support different interaction scenarios.
    \item We provide multi-language user APIs: C++ and C APIs to interface model runners and serializers with compilers and Python APIs to interface inter-process model runners with ML frameworks.     
    \item We show that our library is easy to integrate with \textit{three} different compilers spanning different representations, and carry out extensive evaluations on \textit{four} ML-enabled optimizations on \textit{two} versions of LLVM (V10, V17). 
    \item  We characterize the impact of each communication and serialization options on compilation and training times and analyze additional dependencies and other overheads.
\end{itemize}

\section{Background}
\label{sec:background}

\begin{figure}[h!tb]
    \centering
    \includegraphics[scale=0.4]{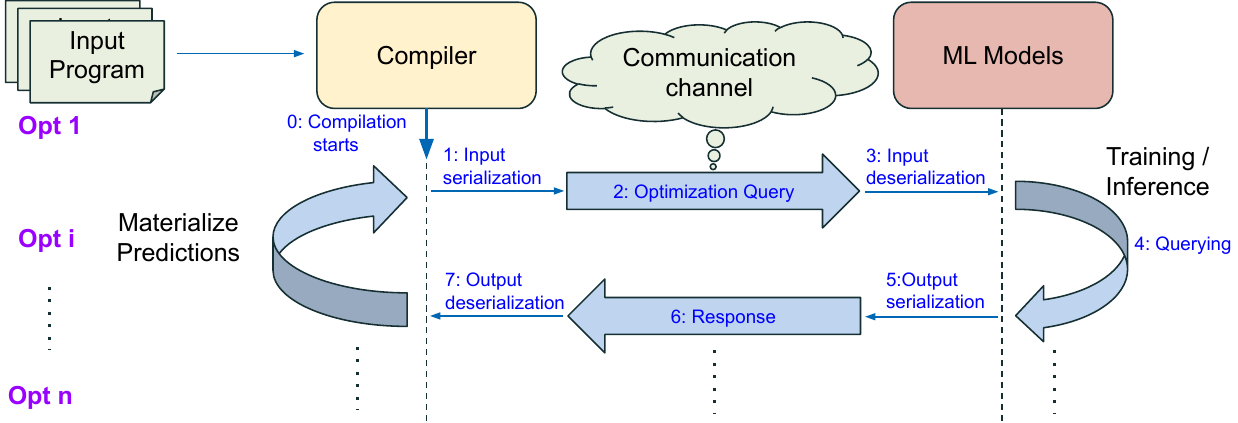}
    \caption{ML-enabled compiler optimizations: (1) Inputs and other metadata required by the model are prepared in the appropriate format. 
    (2) Serialized input is passed on to the model by a suitable communication channel. 
    (3) Input is deserialized to appropriate format. 
    (4) The model is queried to obtain optimization decisions as output. 
    (5) Output is serialized, and (6) Sent back to the compiler optimization as a response. 
    (7) The received response is deserialized, and optimization decisions are taken according to the output. }
    \label{fig:ml-comp-opt}
\end{figure}

\paragraph{ML-enabled Compiler Optimizations}
The process of supporting or fully implementing optimization decisions with one or more ML models involves the steps shown in Fig.~\ref{fig:ml-comp-opt}.
This process repeats until the end of the compilation process for each ML-based optimization.
The above scheme is generic enough to capture any optimization involving single or multiple ML models with multiple two-way interactions. For the cases that would need multiple interactions, steps (1)--(7) are repeated until the final outcome.

More broadly, there are three actors involved in developing and using such an ML-enabled compiler. 
(i) The Compiler expert who develops the compiler optimization, 
(ii) The ML expert who designs the ML model for the optimization problem, and
(iii) The end-user who uses the compiler.
Ideally, compiler experts should use the ML models with minimal understanding of the internals/process specific to ML modeling and the framework on which the model is built to arrive at the result.
Similarly, ML experts should instead design the models with minimal or no understanding of compiler internals, infrastructural details, and integration points, focusing on the optimization objectives and information flow.
For the end-user, however, the presence of ML-compiler optimization should be \textit{transparent}, and \textit{indistinguishable} from the conventional (non-ML based) compilation process.
To achieve this scheme of abstraction/segregation among all three actors, it is important to distinguish between the training and inference flows.

\paragraph{Training}
Typically, training the ML model becomes part of compiler development and build-up, and inference becomes part of the compiler deployment and execution.
However, occasionally this boundary may shift towards the user, like domain-specific training or fine-tuning at deployment time.
Since ML developers usually prefer developing models within a Python-based framework, the training process involving a C++ compiler infrastructure like LLVM requires a communication channel, typically inter-process, while catering to the needs of (de-)serializing data between the native types of C++ and Python. 
The distributed nature of training processes may also require extending communication beyond a single operating system node.

\paragraph{Inference}
When focusing on inference/deployment, compile time and ease-of-use become crucial factors.
The communication and serialization methods involved should take this into account, 
along with considering converting the Python model to a streamlined C++ implementation.
These factors are true even for the simplest forms of communication, like one-time evaluations of the ML model and communicating via flags.
Making the flow transparent to the user also requires a deeper, end-to-end integration with the compiler.

There is no tool providing the necessary layers of abstraction between the three actors while supporting the required training and inference scenarios, not to mention ML-framework independence.
\emph{Designing such a library and evaluating its suitability for diverse use cases} is the challenge we tackle in this paper.

\section{\toolname{}}
\label{sec:ml-compiler-lib}

\begin{figure}[h!tb]
    \hskip-1.1\columnwidth\rlap{
    \includegraphics[scale=0.55]{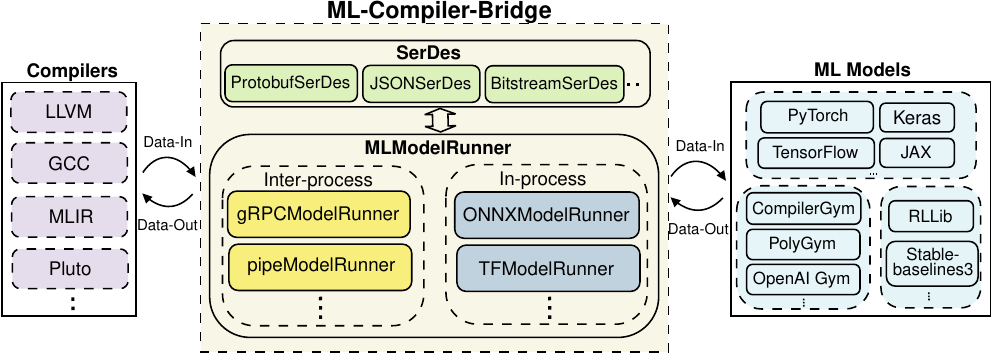}}
    \caption{The compiler instantiates a model runner and sets the input features to be used by the model. \texttt{MLModelRunner} internally invokes
    \texttt{SerDes} to serialize the data in one of the supported formats and query the model. The returned decision is deserialized and provided to the optimization.}
    \label{fig:ml-compiler-lib}
\end{figure}

We propose an abstraction mechanism made of two main components: Serializer and Model Runner.
The \serdes module (de-)serializes the data to/from the requested format, and the \mlRunner module is responsible for communication with the model.
The model runner obtains the serialized data, writes it to a communication channel, queries the model, and deserializes the output received from the model. \toolname{} exposes methods to be invoked by the user to interact with the model decoupled from serialization and communication.
We provide three framework-independent model runners, \grpc, named-pipes, and \onnx, and one framework-specific \tensorflow model runner. 
These can be combined with three different serializations: \protobuf, \json, and bitstream.
The modular design enables new forms of communication and serialization to be added by overriding a minimal set of methods.
Fig.~\ref{fig:ml-compiler-lib} shows the components and interactions of \toolname{}.

\subsection{ML Model Runners}
\label{sec:model-runner}

\begin{figure}[tb]
\begin{lstlisting}[language=C++, style=protobuf, caption={Skeleton of \texttt{MLModelRunner} class}, label={lst:ml-model-runner}]
class MLModelRunner {
public:
  // Exposed to the user; returns model's output
  template <typename T> T evaluate() {
    return *reinterpret_cast<T *>(evaluateUntyped());
  }
protected:
  // To be overridden by derived classes
  virtual void *evaluateUntyped() = 0;
};
\end{lstlisting}
\vspace{-\baselineskip}
\end{figure}

We provide two classes of model runners.
The inter-process class provides the easiest mechanism to decouple Python models from a compiler running as a separate process. 
The in-process class assumes that the \mlmodel is readily available in a compiled form 
and can be accessed within the compiler through a specific API.
Clearly, in-process communication is designed with inference and deployment in mind, while inter-process communication enjoys more diverse use cases.
Model runners may support simple ML queries and feed-forward networks as well as more involved Reinforcement Learning (RL) algorithms or Graph Neural Networks (GNNs).

Internally, \mlRunner is the abstract base class from which the other model runners are derived (List.~\ref{lst:ml-model-runner}). 
It exposes two APIs: \texttt{populateFeatures()} populates the input features, and 
\texttt{evaluate()} queries the model. 
The latter returns the output of the model and is templated according to the expected output type. 
Internally, \texttt{evaluate()} invokes \texttt{evaluateUntyped()} that is to be overridden by the concrete model runner classes that derive from \mlRunner. 
The \mlRunner interfaces with the methods of \serdes using the \texttt{populateFeatures()} so as to serialize the inputs.
The method \texttt{populateFeatures()} is implemented as a variadic function that uses variable-length key-value pairs as arguments. 
The key is a string identifier that describes the input, and the value is of template type.

\subsubsection{Inter-process Model Runners}

\grpcRunner uses \grpc{} may run the model and compiler on different machines, and \pipeRunner uses \emph{named pipes} for single-machine scenarios only.
At training time, the compiler acts as a server and the Python-based ML model acts as a client.
The sequence of steps is described as follows:
\begin{enumerate}[label=(\arabic*)]
    \item Compilation starts and the compiler \textit{listen}s for queries at the \texttt{wait()} call inserted at the point of interest.
    \item The Python model starts training; this can be started concurrently with Step~(1).
    \item When input from the compiler is required, the model sends requests to the compiler with appropriate queries and waits for the response.
    \item The compiler gets out of the blocked state and processes the query to generate an appropriate response.
    \item The response is sent back to the client, and the model goes on to completing training on that input.
\end{enumerate}

Inference follows the same steps, yet the compiler becomes the client and the model becomes the server so as to support a regular compilation process.

\begin{figure}[t]
\begin{lstlisting}[frame=tb, language=protobuf2, style=protobuf,caption={Example gRPC function declaration}, label={lst:rpc-service}]
service gRPCExample{
  // RPC function for training integration
  rpc getObservation(Action) returns (Observation) {}
    
  // Mandatory RPC function for model evaluation
  // Blocking call; Released upon receiving a response
  rpc getAdvice(Observation) returns (Action) {}
}
\end{lstlisting}
\vspace{-\baselineskip}
\end{figure}

\paragraph{\grpc Model Runner}
\label{sec:grpcModelRunner}

\grpc~\cite{grpcWang1993} provides RPC methods specifying the type of input and output in \protobuf format~\cite{protobuf}.
During the build process of the library, the proto files are automatically translated to C++ and Python code by invoking the \texttt{protoc} compiler.
An example is shown in List.~\ref{lst:rpc-service}.
The generated code defines the \texttt{Service} class that exposes the RPC methods to be overridden by the user in the optimization that makes use of \grpcRunner.
Due to design constraints of \grpc, we only support \protobuf serialization with \grpcRunner.

\grpcRunner takes in the server address and the port number at which the connection is to be established.
In training mode, \grpcRunner starts the server and starts listening for an RPC call invoked by the model. The overridden RPC method is directly called by the Python model to generate new observations by applying the action predicted by the model.
In inference mode, \grpcRunner starts the \grpc connection at the given address and port.
\texttt{evaluateUntyped()} is overridden to invoke the RPC method defined by the Python model
after preparing the input data, and\texttt{getAdvice()} serves as the RPC method for inference. 

\paragraph{Pipe Model Runner}
As the name suggests, \pipeRunner relies on named pipes for inter-process communication (the \texttt{mkfifo} system call).
Pipes provide a simple and effective means of communication that is local to the machine without any network or security constraints.

As pipes are unidirectional, the \pipeRunner creates the read and write pipes for communication.
The read pipe in the compiler obtains the data written by the model in Python, and the write pipe
provides the data into the pipe that is read by the model on the other end. The \texttt{evaluateUntyped} method is overridden to read and write into the pipe appropriately. \texttt{read()} is a blocking call 
forcing the compiler to wait till data is written by the model. Once the data is written, the model gets to a blocking state by invoking \texttt{read()} on the second pipe waiting for the response from the compiler.
The pipe model runner ensures proper opening, closing, and clean up. 
\pipeRunner provides a simpler interface for establishing communication as the user directly invokes \texttt{evaluate()} after setting the inputs.

\subsubsection{In-process Model Runners}
In-process model runners are designed to provide an effective means of compiler deployment.
It is important to optimize the inference time as it adds up to the overall compile time.
One may obtain significantly lower compile times by removing inter-process communication overhead, and by turning the complications of a compiled model into an advantage, by reducing the query time compared to models running in Python. Serialization/deserialization overhead is also lowered.

\paragraph{\onnx Model Runner}
The Open Neural Network Exchange~\cite{onnx} (ONNX) is an open format to represent machine learning models.
Models built from various frameworks like \tensorflow, PyTorch, etc. can be represented in \onnx format in an interoperable manner.
Additionally, it supports a wide variety of hardware architectures ranging from edge devices to general-purpose CPUs and GPUs. 
Once the model is trained in Python, it is converted into a common \onnx representation and is imported into the compiler via the \onnx runtime. 
\onnxRunner exposes the necessary wrapper APIs to read the \onnx model, query it with inputs and obtain outputs. \onnxRunner also RL models.

\begin{figure}[h!tb]
    \centering
    \includegraphics[scale=0.53]{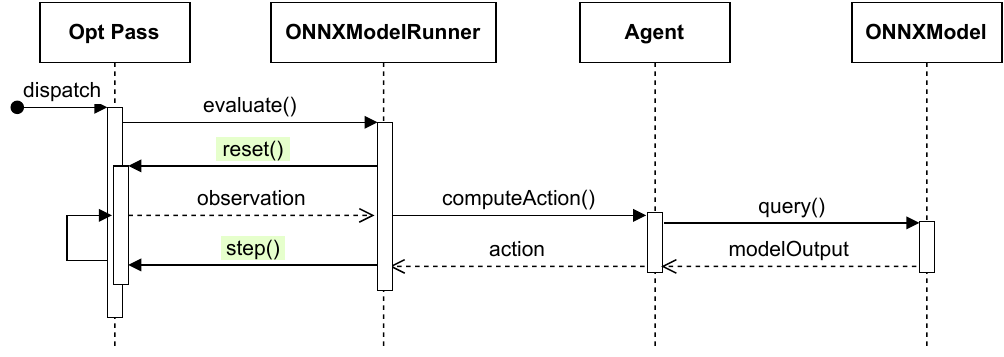}
    \caption{Sequence diagram indicating different events and the interaction between various classes for RL based optimization by \onnxRunner. Only the methods that highlighted are to be \colorbox{lgreen}{overridden} by the user. Other methods are internal to the library.}
    \label{fig:onnx-sequence}
\end{figure}

\paragraph{\onnxRunner for RL}
For RL, the agent is usually the learner trained to predict appropriate actions given the observations from the environment.
Exporting a trained model to \onnx implies exporting only the agent. 
To facilitate RL-based interaction for a generic multi-agent scenario between the environment and the agents, \onnxRunner provides Environment and Agent classes separately and accesses the APIs internally.
The sequence of events describing this interaction is shown in Fig.~\ref{fig:onnx-sequence}. 

\onnxRunner exposes the \texttt{Environment} class with APIs for standard \texttt{step} and \texttt{reset} operations along with \texttt{setDone()} API to indicate the end of the episode. 
\texttt{step()} returns the next observation given an action. Internally, the step operator applies the action predicted by the agent (model) to \textit{move on} to the next state and returns the new observation from the environment. \texttt{step()} also signals if the terminal state is reached by invoking \texttt{setDone()} to stop the current prediction. \texttt{reset()} resets the environment to the initial state and returns the initial observation. Hence \onnxRunner involves the Reset operator first to obtain the initial observation. This sequence of APIs is invoked within the \texttt{evaluateUntyped()} of \onnxRunner and is shown in the Listing~\ref{lst:onnx-evaluate}.
The optimization pass using the \onnxRunner should inherit from \texttt{Environment} and override \texttt{step()} and \texttt{reset()} depending on the optimization requirements.

\onnxRunner queries the model using the C++ APIs. A map containing the identifier of the agent (label) and the corresponding model path is passed while instantiating the \onnxRunner. In the case of multiple agents, the identifier of the next one to use is set by the \texttt{Environment} while returning the observation. \onnxRunner queries the corresponding agent 
with the observation to obtain the requested action. This process goes on untill \texttt{Environment} invokes \texttt{setDone()}.

\begin{figure}[t]
\begin{lstlisting}[frame=tb, language=C++, style=protobuf, caption={Snippet from \onnxRunner showing the environment-agent interaction to generate an observation}, label={lst:onnx-evaluate}]
void *ONNXModelRunner::evaluateUntyped() {
  Observation obs = this->env->reset();
  while (true) {
    Action action;
    // current agent
    auto current_agent = this->agents[this->env->getNextAgent()];
    action = current_agent->computeAction(obs);
    obs = this->env->step(action);
    if (this->env->checkDone())
      break;
  }
  return nullptr;
}
\end{lstlisting}
\vspace{-\baselineskip}
\end{figure}

\paragraph{\onnxRunner for plain ML models}
\onnxRunner can also be used to query non-RL models by directly invoking the \texttt{evaluate} method upon instantiating the object with the path to the ONNX model.

\paragraph{\tensorflow Model Runners}
This is a framework-specific model runner built on the \tensorflow ahead-of-time (AOT) saved model.
There are two implementations: (i) \textit{``Release Mode Model Runner''} 
used in production environments,
(ii) \textit{``Model Under Training Model Runner''}
intended either for finetuning or when quickly evaluating candidate models and parameters. 
TFLite is a scaled down \tensorflow interpreter designed to be embedded in native binaries, and can be used to further reduce overheads.

The \tf model runner uses the AOT saved model compiler
which produces a header exposing the model as a C++ class, and a native object file with its implementation. 
The model runner reduces again to a simple adapter~\cite{gamma1995design} around that class.
The compiler binary does not expose new runtime dependencies as it is statically linked, and this simplifies its deployment. 
Note that the model compiler can be configured to generate code loading the weights from a file passed via the command line to the LLVM compiler.

\begin{figure}[h!tb]
    \centering
    \includegraphics[scale=0.37]{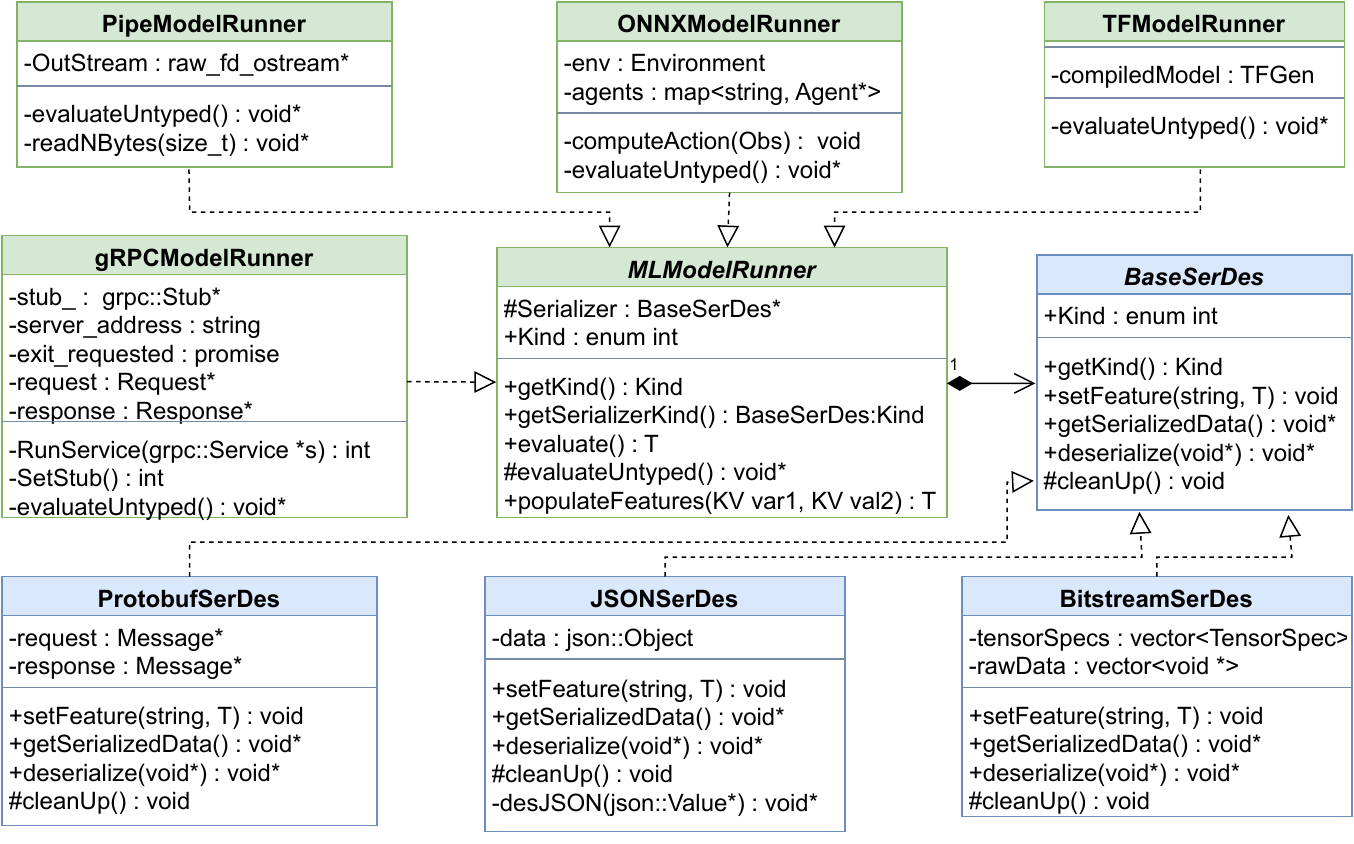}
    \caption{Class diagram of \toolname}
    \label{fig:class-diagram}
\end{figure}
\subsection{SerDes: Serializer and Deserializer Module}
\label{sec:serializers}

When data is transferred, specifically across two processes, it is important to convert data that is present in the native types (of C++ and Python) from one format to another. 
This is the purpose of (de-)serialization as implemented by the \serdes module.

Internally, the \mlRunner interacts with \serdes to (de-)serialize C++ native data to model-specific types and back.
The choice of (de-)serialization depends on the optimization and ML model. 
We currently provide three options: bitstream, \json, and \protobuf.
They vary in terms of usage scenario, usage effort, and (de)serialization time. 
\serdesntt effectively abstracts away the underlying mechanism while providing the flexibility of different serialization options.

\paragraph{Base SerDes}
Internally, each \serdesntt is derived from the \baseSerDes class. 
\serdes uses key-value based serialization as described in Sec.~\ref{sec:model-runner}. 
The \texttt{populateFeatures} method of \mlRunner invokes the appropriate version of the overloaded \texttt{setFeature()} exposed by \baseSerDes to serialize inputs.
These methods are overridden by the \serdesntt classes that derive from \baseSerDes according to the underlying serializer. 
This class also exposes the \texttt{deserialize} method to deserialize the received data and is overridden by the derived classes to obtain the data in native types. 
Our library supports (de)serializing in basic (int, float, double, string, bool) and compound (vector, list) data types.

\paragraph{\protobuf{} \serdesntt}
\protobuf{} \serdesntt needs the user to provide the input and output data specifications in a proto file. These are compiled to generate the C++ and Python sources (Sec.~\ref{sec:grpcModelRunner}). 
\protoSerDes serializes the input key-value pair by overriding the \texttt{setFeature} methods to set the appropriate fields of the \textit{message} described in the proto file.
Deserializing protobuf data to the native format only involves reading and returning the appropriate fields of the \textit{message}. 
Except for providing the proto file, \protoSerDes is transparent to the user.

\paragraph{\json{} \serdesntt}
\jsonSerDes overrides the \texttt{setFeature} methods to populate the \json buffer appropriately, given the key-value pairs. 
Similarly, the received data is deserialized by first converting it to a \json object, then the JSON fields are casted to native types and returned.
\json{} \serdesntt is also transparent to the user.

\paragraph{\bitstream{} \serdesntt}
The bitstream starts with a \json header which specifies the key (identifier), type and shape of the tensors, and the order in which they will be serialized.
Tensor values themselves are dumped as raw bytes.
The received bitstream is interpreted based on the type and shape specified in the header and converted to native types. 
Processing the header induces negligible overhead if communicated data does not involve complex data types.
Internally, \bitSerDes overrides the \texttt{setFeature} methods similar to the other \serdesntt to expose the functionality. Fig.~\ref{fig:class-diagram} shows the class diagrams~\cite{gamma1995design} of model runners and \serdesntt.

\subsection{C-APIs}

We provide C wrappers around the C++ implementation to integrate with C-based compilers. 
These wrappers are C++ files written in C-style. Each method internally queries the original C++ implementation and returns results in a way compatible with C calling conventions.
This code is built as a separate library that may be linked with a C-based compiler. We used it with the Pluto polyhedral compiler in particular.

\begin{figure}[h!tb]
\begin{lstlisting}[frame=tb, language=C++, style=protobuf, caption={C++ APIs of \toolname{}}, label={lst:user-view-api}]
#include "MLCompilerBridge/MLModelRunner.h"
#include "MLCompilerBridge/yourMLModelRunner.h"

// Instantiate the required model runner with SerDes type
MLModelRunner *MLRunner = std::make_unique<yourModelRunner>(Arg, SerDes::Kind::yourSerDesType);
// Process Input Features
std::pair<std::string, InType> p = ... // Input
MLRunner->populateFeatures(p);
// Get ML Advice/Output
OutType advice = MLRunner->evaluate<OutType>();
// Use the obtained advice
...
\end{lstlisting}

\end{figure}

\subsection{Extensions}
Both \mlRunners and \serdes can be easily extended to support new model runners and serializers. New runners may include TVM~\cite{tvm2018OSDI}, ahead-of-time compiled PyTorch models and FlatBuffers~\cite{flatbuffers}, and serialization also supports YAML formats.
New model runners can be contributed by inheriting \mlRunner and overriding the \texttt{evaluateUntyped} method according to the model runner.
Similarly, a new (de)serializer can be added by inheriting \baseSerDes and overriding the \texttt{setFeature} and \texttt{deserialize} methods specific to the new serializer.

\subsection{Error Checking and Recovery}
\label{sec:errors}

The model runners and \serdesntt modules are designed to handle compiler/model crashes, communication failures, and infinite loops.
The failures are handled appropriately by allowing graceful termination of the processes. In the case of \grpc, we have implemented an exponential backoff algorithm to attempt retries to overcome the failures due to the delays in communication resulting from any network-related issues and packet losses. The communication fails gracefully upon exhausting the number of retries. In all other cases, we use a timeout based mechanism for handling the failure. These mechanisms proved invaluable in practical experiments due to compiler bugs and network errors.

\subsection{Compiler/ML Experts View}

To use \toolname{}, developers need to invoke a minimal set of APIs by instantiating the necessary model runner with appropriate options specifying the \serdesntt type. List.~\ref{lst:user-view-api} illustrates this on an example of invoking a user-defined model runner with a user-defined \serdesntt from the compiler. A similar API abstracting the communication and \serdesntt in Python is provided (List.~\ref{lst:user-view-pyapi}) to query the ML model with inter-process model runners and respond back. 

\begin{figure}[h!tb]

\begin{lstlisting}[frame=tb, language=Python, style=protobuf, caption={Python APIs of \toolname{}}, label={lst:user-view-pyapi}]
import CompilerInterface as CI

# Instantiate the required CompilerInterface with serdes type
interface = CI.YourCompilerInterface(Arg, yourSerdesType)
while True: 
    # Send buffer data to compiler and wait for next request
    request = interface.evaluate()
    # Query model to get advice
    # Populates buffer with advice (serialized and stored in serdes)
    interface.populate_buffer(advice)
    # Break on condition
\end{lstlisting}
\end{figure}

\section{Use Cases: ML-LLVM optimizations}
\label{sec:ml-llvm-project}

We integrated \toolname{} with four ML-based compiler optimizations in LLVM: phase ordering~\cite{Jain-POSET-RL-2022}, loop distribution~\cite{shalini-rl-loop-distribution-2022}, register allocation~\cite{VenkataKeerthy-2023-RL4ReAl} and method inliner~\cite{trofin20MLGO}. The first three optimizations are built using RLLib~\cite{liang18bRllib} with PyTorch~\cite{pytorchNeurIPS019} and LLVM V10, using program embeddings called IR2Vec~\cite{VenkataKeerthy-2020-IR2Vec}. The fourth optimization---inlining---uses TensorFlow~\cite{tensorflow2015-whitepaper}, is built within LLVM V17, and uses feature-based representations~\cite{trofin20MLGO}. There are two ML based register allocators~\cite{VenkataKeerthy-2023-RL4ReAl, mlgo-regalloc-rfc} available for LLVM; 
we chose the former because it emphasizes finer-grained, high-bandwidth interactions with an ML model.
All the components are configured, compiled and linked during the regular build process of LLVM.
Integration challenges range from redesigning the entire framework of the original publication, to minor changes to the communication mechanisms.

\subsection{Phase Ordering of Optimization Passes}
\label{sec:posetrl}
\posetrl predicts the ordering sequence of passes to jointly optimize code size along with execution time. 
An RL agent is trained with the DDQN algorithm~\cite{van2016deep-ddqn} to predict a subsequence as action, given program embeddings as input observation. There are about $15$ predetermined subsequences provided by the authors.
The predicted optimization subsequence is applied on the input program, and the embeddings corresponding to the transformed program are used as the new observation. This process goes on until reaching a threshold on the number of subsequences.

In the published version, the above process was not integrated within LLVM but driven from a Python model. 
An LLVM-opt process was spawned, passing the optimization sequence through a compiler flag for each prediction by the agent.
In addition, embeddings involve spawning yet another process to invoke IR2Vec on the \texttt{.ll} IR file generated by the compiler. 
A similar strategy was in place for training.

We revisited the above using \toolname{} to operate directly within LLVM as a new transformation pass. Our new \posetrlpass implements a pass manager that applies the predicted optimization sequence, and also generates the next observation by invoking IR2Vec.
The \mlRunner communicates with the model and serializes the data to be transferred.
The model communicates the predicted optimization subsequence as an integer ID (one among $15$) to \posetrlpass, and the $\mathbb{R}^{300}$ module-level embedding vectors are sent to the model for the next prediction. 
Integrating with the \onnx{} model runner only amounts to extending the \texttt{Environment} class and overriding the \texttt{step}, \texttt{reset} methods. We also override \texttt{setDone()} to signal the end of the episode upon reaching the threshold.

\subsection{Loop Distribution for Vectorization and Locality}
\label{sec:loopdist}
Jain et al.~\cite{shalini-rl-loop-distribution-2022} improve loop distribution by modeling SIMD parallelization and locality optimization opportunities. 
It uses two RL agents with fully-connected networks to identify the vertex processing order and when to distribute. Along with these agents, a Gated Graph Neural Network (GGNN)~\cite{GGNN} processes the connected components of the dependence graph, where each node holds the embeddings for the corresponding instructions.

During training, a Python driver spawns a process to invoke the Loop Distribution pass. The RL model processes the input graph and predicts the sequence of instructions to be packed together as a loop. Upon applying the prediction, the rewards indicate the effectiveness of distribution. All these steps involve model-compiler interaction via file I/O.
Inference itself is integrated with LLVM using Python wrappers.

In this paper, we eliminate the need for Python wrappers, file I/O and and spawning new processes. The model runners internally (de-)serialize data depending on the chosen \serdes and the \mlRunner. For the runners that use serialization, the input graph is represented as key-value pairs, and a variable length matrix in $\mathbb{R}^{n \times 300}$ encodes the sequence of $n$ $300$-D instruction embeddings. The output takes the form a variable-length integer array with node identifiers that are to be distributed.

\subsection{RL-Based Register Allocation}
\label{sec:rl4real}
We also evaluate \rlforreal, an RL-based register allocator implementing the splitting, coloring, and spilling sub-tasks as separate RL agents on LLVM's Machine IR.
These RL agents pose a formidable engineering challenge in interfacing the model with the compiler during both training and inference.
Unlike other optimizations that need one single communication at the end, \rlforreal involves \emph{multiple interleaved communications rounds} to obtain a new observation and let the relevant agent make the next prediction. 
Also them RL agents are arranged hierarchically: the outcome of one agent determines which agent would be invoked next. 
Unlike other use cases, this optimization involves transferring an interference graph where each variable is associated with
a $\mathbb{R}^{n \times 100}$ matrix, and where each one of the $n$ instructions in the live range of the variable is represented in $100$-D, 
a variable-length integer array to specify interferences and use points, and 
a variable-length floating point array of spill weights. 
Other metadata like function name, file name, and status are also sent as string fields. 
The model returns key-value pairs mapping variables to split or color decisions. 
Both training and inference use \grpc and \protobuf serialization.

We will investigate different communication and serialization improvements in this paper, with specialized scenarios for distributed training and deployment-friendly inference.

\subsection{LLVM Inliner}
\label{sec:inliner}

The inliner pass traverses call sites in a bottom-up fashion, one connected component of functions at a time.
For a given component a working queue is initialized with the set of all static call sites.
As the algorithm marks some call sites for inlining, it appends the former callee's call sites to the work queue. 
The decision to inline or not is made in two steps. 
First, it determines legality and whether the user provided any guidance (always/never inline). 
Only if the operation is legal and non-mandatory, a heuristic determines its profitability. 

The decision is driven by a simple RL based model.
It takes a number of scalar features characterizing both the caller/callee (instruction counts, basic block counts, maximum loop depth), the call site itself (the number of compile-time constant parameters), as well as module-wide features (the current number of functions and statically known call edges). 
For the published version~\cite{trofin20MLGO}, the cost metric was size, with no reliance on dynamic profile data.
The implementation uses AOT compiled \tf model for inference with C++ APIs.
We modularized it to use any model runner.

\section{Evaluation}
\label{sec:evaluation}
We measure compilation time on an Intel Xeon SkyLake W2133 with 6 cores, 12 threads and 32GB RAM. Training time is measured on an Intel Xeon W1390P with 8 cores, 16 threads, 64GB RAM and an Nvidia 3060 GPU.
We evaluate \posetrl, \loopdist and \rlforreal with \grpc{}, \pipe and \onnx model runners and different \serdesntt options, and take the median of 3 runs.
Most experiments use \specI and \specII benchmarks.

\subsection{Impact on Deployment}
Tab.~\ref{tab:compile-times} shows the \posetrl compile time using different model runners.
Within the in-process runners, we use \onnx for PyTorch models and RLLib.
Overall, in-process runners achieve better compile times in all cases in comparison with any of the inter-process ones.
Among the latter, \grpc{} has higher compile times ($6.8$--$7.6\%$) compared to pipes, with \json and bitstream \serdesntt. This is because of the overheads associated with establishing connections and invoking RPC methods. Pipes with \bitstream \serdesntt yield slightly higher performance than \json \serdesntt due to the lower (de-)serialization overhead with bit streams.
\onnxRunner yields a $7.2 \times$ speedup with \posetrl compared to the original method in Sec.~\ref{sec:posetrl} that involved spawning new processes to invoke the compiler and other dependencies.

In-process model runners natively support multithreaded compilation, while inter-process model runners necessitate concurrently running multiple instances of the model resulting in a high memory and compute overhead. Tab.~\ref{tab:multithreaded-compile-times} shows compile times with in-process model runners on LLVM Inliner and \rlforreal optimizations by varying the degree of parallelism.
As LLVM Inliner and \rlforreal respectively rely on \tf and PyTorch (and RLlib), we use \tf and \onnx model runners accordingly. In comparison to the original \grpc based inference flow of \rlforreal, the ONNX runner reduces compile time by $22.4 \times$ and $19 \times$ using $8$ threads and $1$ thread respectively.
Using \rlforreal results in a higher compile time, as it involves a larger number of model-compiler interactions.
This overhead is effectively reduced by using the model runners exposed by \toolname{}.

Similar trends are observed for RL-driven loop distribution~\cite{shalini-rl-loop-distribution-2022} on TSVC~\cite{tsvc} and the LLVM Test Suite~\cite{llvm-test-suite}. The ONNX model runner yields an improvement of $16 \times$ in comparison to the original Python wrapper.
\begin{table}
\centering
\caption{Compile time (in seconds) for \posetrl.}
\label{tab:compile-times}
\resizebox{\columnwidth}{!}{%
\begin{tabular}{lrrrrr}
\toprule
 & \multicolumn{1}{l}{\textbf{Original}} & \multicolumn{1}{l}{\textbf{gRPC}} & \multicolumn{1}{l}{\textbf{Pipe + JSON}} & \multicolumn{1}{l}{\textbf{Pipe + Bitstream}} & \multicolumn{1}{l}{\textbf{ONNX}} \\
 \midrule
\textbf{SPEC06} & 5,829 & 1,318 & 1,236 & 1,227 & 1,140 \\
\textbf{SPEC17} & 10,342 & 1,221 & 1,141 & 1,132 & 1,093 \\
\bottomrule
\end{tabular}%
}

\medskip
\centering
\caption{Multithreaded compile time with -O3 (in s) with in-process model runners. Compile time with \grpc is shown for \rlforreal for comparison.}
\label{tab:multithreaded-compile-times}
\resizebox{\columnwidth}{!}{%
\begin{tabular}{llllll}
\toprule
 & \textbf{gRPC} & \multicolumn{1}{c}{\textbf{1 Thread}} & \multicolumn{1}{c}{\textbf{2 Threads}} & \multicolumn{1}{c}{\textbf{4 Threads}} & \multicolumn{1}{c}{\textbf{8 Threads}} \\
\midrule
\textbf{\begin{tabular}[c]{@{}l@{}}LLVM Inliner \\ (TF Runner)\end{tabular}} & \multicolumn{1}{c}{-} & \multicolumn{1}{r}{596} & \multicolumn{1}{r}{501} & \multicolumn{1}{r}{361} & \multicolumn{1}{r}{307} \\
\textbf{\begin{tabular}[c]{@{}l@{}}RL4ReAl\\ (ONNX Runner)\end{tabular}} & \multicolumn{1}{r}{5,572} & \multicolumn{1}{r}{291} & \multicolumn{1}{r}{257} & \multicolumn{1}{r}{248} & \multicolumn{1}{r}{248} \\
\bottomrule
\end{tabular}%
}
\vskip-1\baselineskip
\end{table}

\subsection{Impact on Training}
In this section, we evaluate the effectiveness of \toolname{} during the training of \posetrl and \rlforreal. We use inter-process model runners for training.

\subsubsection{Training Time}
Fig.~\ref{fig:training-time-microbenchmarking}(a) shows the cumulative training time and number of training iterations observed in \posetrl.
We obtain \emph{large improvements} in the training time \emph{across all the model runners}.
We see similar trends with \grpc{} and \pipe, as explained in the previous experiment.

The original training process of \posetrl involves spawning processes that takes $\approx 10$Ks to complete $500$ iterations. 
In comparison, the \grpc model takes about $5.7$Ks, while the pipes with JSON and bitstream serialization options take about $5.5$Ks each.
Throughout the iterations, we observe an overhead of about $20$s between JSON and bitstream serialization options.
This minimal overhead is associated with the additional serialization effort involved while using JSON \serdesntt.
However, using the inter-process model runners enables an \textit{end-to-end integration of model and the compiler} while training yields a significant improvement.

\subsubsection{Multi-Worker Support}
\begin{figure*}[h!t]
    \centering
    \subfloat[\centering{\footnotesize Training times of \posetrl with different Model Runners}]{{\includegraphics[scale=0.27]{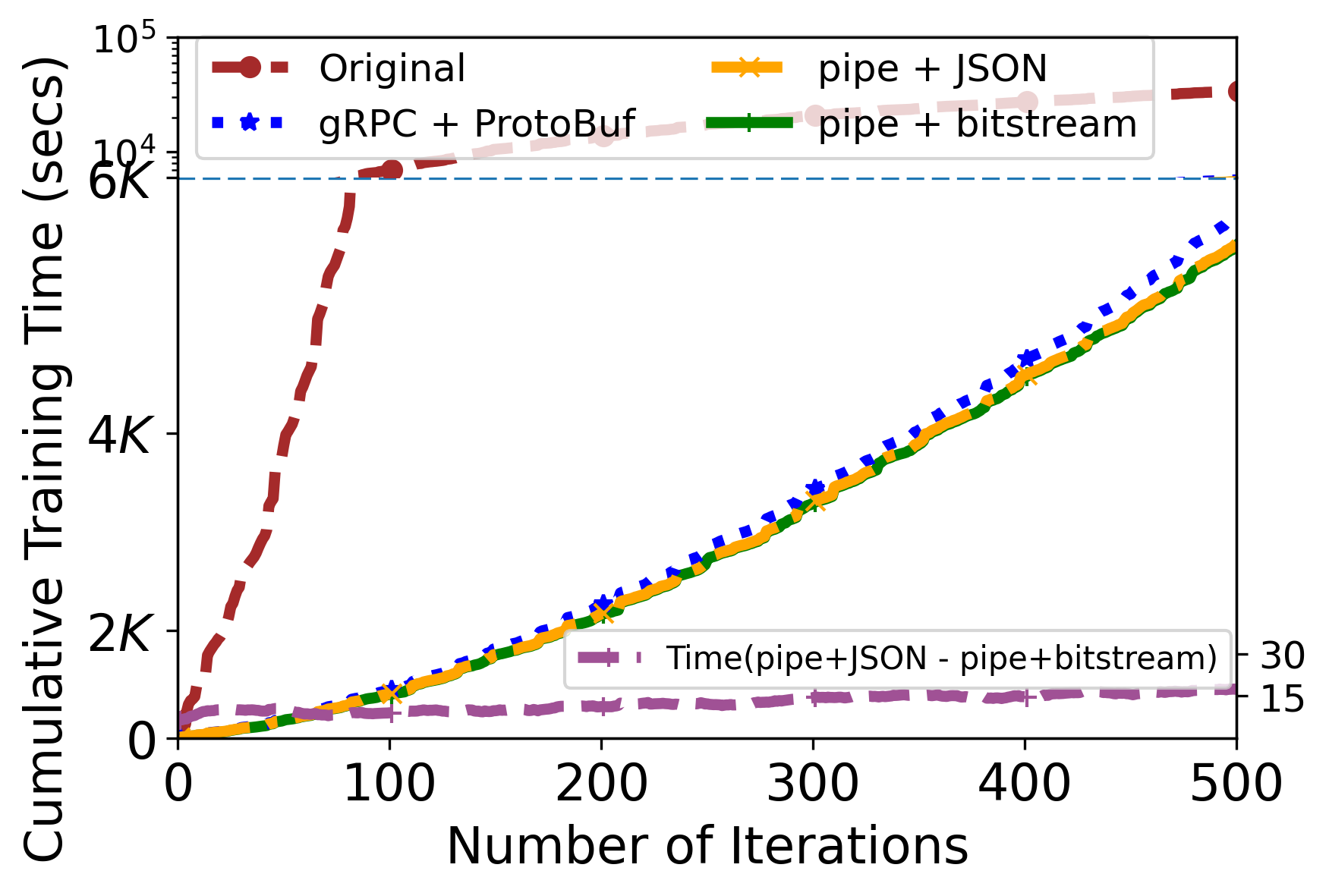}}}
    \subfloat[\centering{Training times of \rlforreal with CPU/GPU multi-workers}]{{\includegraphics[scale=0.223]{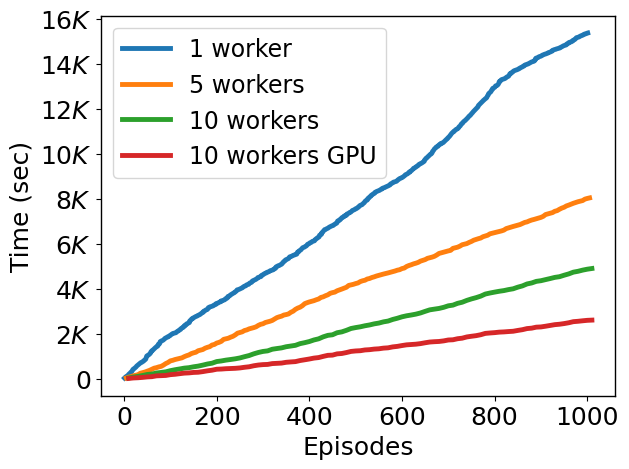}}}
    \subfloat[\centering{Microbenchmarking of individual Model Runners}]{{\includegraphics[scale=0.223]{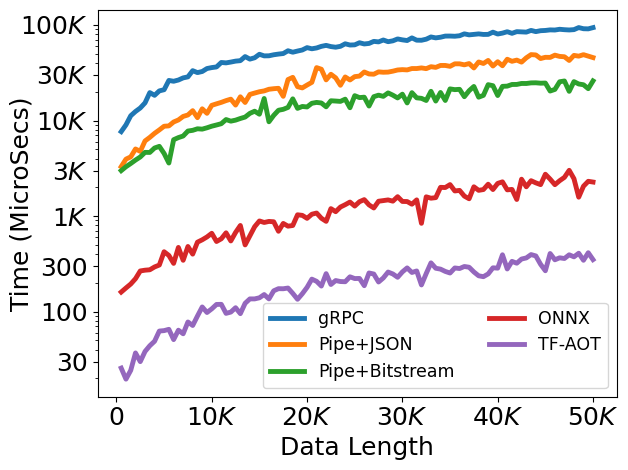}}}
    \subfloat[\centering{MLIR performance}]{{\includegraphics[scale=0.223]{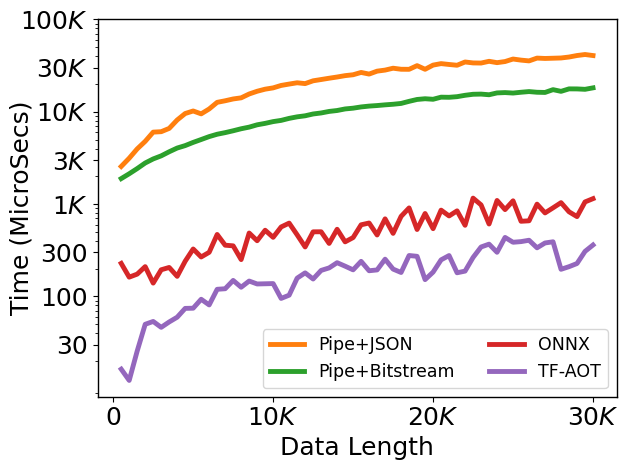}}}
    \subfloat[\centering{Pluto performance}]{{\includegraphics[scale=0.223]{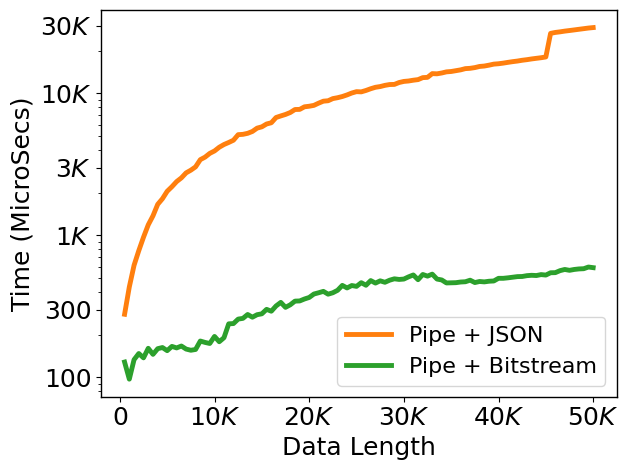}}}
    \caption{Performance characterization of model runners on different compilers and optimizations}
    \label{fig:training-time-microbenchmarking}
\end{figure*}

\toolname{} supports multi-worker training on both CPUs and GPUs. 
To support multiple workers while using \grpc{}, we expose a method taking an array of ports to establish connections with each worker. Similarly, multi-worker support with pipes is enabled by instantiating one pair of pipe per worker. We extended \rlforreal to handle multi-worker scenarios; training times are shown in Fig.~\ref{fig:training-time-microbenchmarking}(b) for CPU and GPU workers. 
Using $10$ workers with a GPU trainer takes about $2$ seconds per episode, while a CPU trainer with <$10$, $5$, $1$> workers takes <$4$s, $8$s, $15$s> respectively.
We obtained similar trends among the workers even upon using pipes for communication.

\subsubsection{Using Different RL Policies}

One may train and deploy models with different RL policies without impacting the compiler. 
For this experiment, we evaluate \rlforreal with the different RL policies provided by RLlib. We perform hyperparameter tuning using Tune~\cite{liaw2018raytune}.
We trained the models with PPO~\cite{schulman2017ppo}, APPO~\cite{schulman2017ppo}, and A2C~\cite{pmlr-v48-a2c16} policies untill convergence.
On the \specII benchmarks, this resulted in $2\%$ improvement on average using the APPO policy.
The PPO and A2C perform similarly to original paper.

\subsection{Round-Trip Time}
\label{sec:rtt}
Let us finally isolate the Round-Trip Time (RTT) of each model runner as a limit study of the achievable communication throughput.
We consider random floating point vectors of increasing length ranging from $500$ to $50$K elements in steps of $500$.
The model itself is a single fully-connected layer that consumes the vector and returns a scalar float.
Fig.~\ref{fig:training-time-microbenchmarking}(c) shows the RTT of the whole process.
The TF and ONNX runner achieves a very high throughput with a total RTT of $21$ and $68$ms respectively; 
while Pipes+JSON and Pipes+Bitstream yield $3154$ms and $772$ms respectively, and \grpc yields a larger RTT of $5948$ms. 
These differences can be attributed to the serialization and communication overhead.
The TF and ONNX runners benefit from in-process communication, proving to be suitable candidates for deployment. The higher throughput of TF is due to the AOT precompiled model.
The Pipe runner proves to be a good candidate for training on local machines. 
And the \grpc runner provides support for training in a distributed environment. 
This makes \emph{all the model runners important in their own way}.

\subsection{Gym Integration} 
\begin{figure}[h!tb]
    \centering
    \hspace{-0.4cm}
    \includegraphics[scale=0.29]{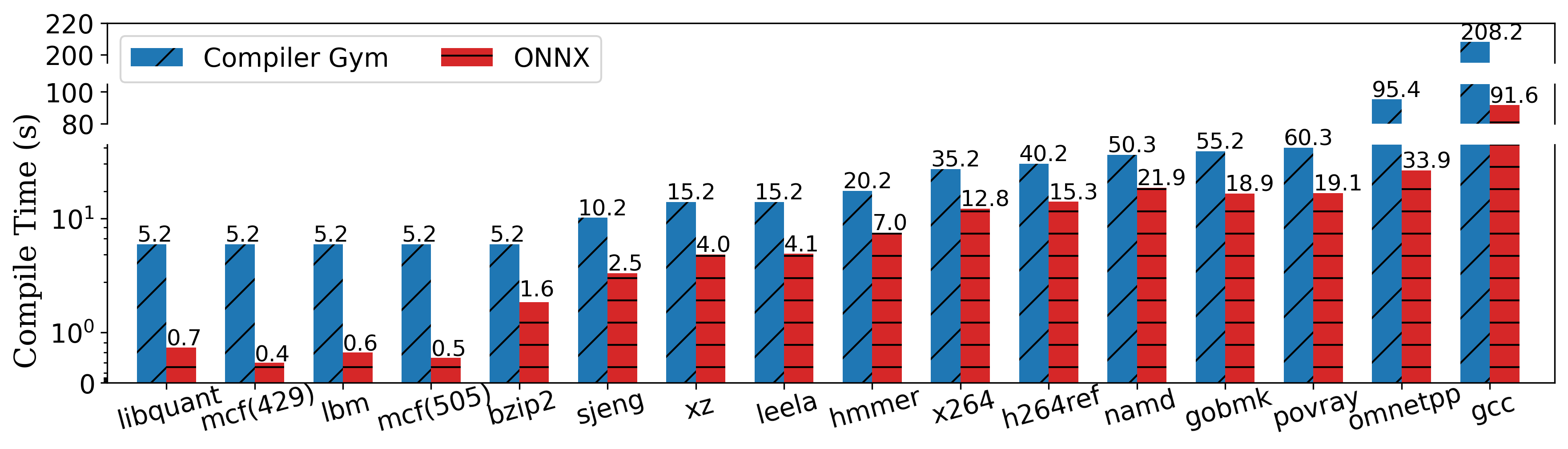}
    \caption{Compile times on using phase ordering for code size model with CompilerGym and ONNX model runner}
    \label{fig:compilergym-comparison}
\end{figure}

We carried out additional experiments to evaluate the benefits of our library in the context of a state of the art RL Gym. The two goals are to facilate deployment and to reduce compilation time by using in-process model runners.
For this purpose, we trained the pass ordering for code size of CompilerGym~\cite{cummins2021compilergym} and
exported the resulting model in the ONNX format. 
We then used our ONNX model runner within LLVM to materialize predictions and generate code. 
The inference times are shown in Fig.~\ref{fig:compilergym-comparison}, with speedups ranging from $2\times$ to $13\times$. 
These are primarily due to gRPC overheads in CompilerGym, as shown in Fig.~\ref{fig:training-time-microbenchmarking}(c).

\subsection{Domain-Specific Compilers}

Given LLVM's dominance in the general-purpose and backend compiler landscape, it forms the natural basis for most ML/RL tools (Tab.~\ref{tab:rel-works-comp}).
However, some ML-based domain-specific optimizations target higher-level frameworks like MLIR~\cite{lattner-mlir-2021} and the polyhedral compilers Pluto~\cite{bondhugula-pluto-2008} and PoCC~\cite{pocc}.
Let us illustrate the cases of MLIR and Pluto.

\paragraph{Integration with MLIR}
Given that end-to-end ML compilers based on MLIR are still undergoing rapid changes \cite{IREE}, we designed a simple experiment to demonstrate the integration of \toolname{} with MLIR. 
We wrote a custom pass in MLIR to communicate data with a dummy ML model to mimic a typical ML-compiler interaction. 
We use the same experimental setup as discussed in Sec.~\ref{sec:rtt} and measure the round-trip time.
The results are shown in Fig.~\ref{fig:training-time-microbenchmarking}(c).
This opens up ML-based optimizations in MLIR-native compilers such as IREE and OpenXLA \cite{IREE}, Triton \cite{Triton}, Polygeist~\cite{moses-polygeist-pact21}, and many other frameworks.

\paragraph{Integration with Pluto}
We also experimented with the polyhedral source-to-source compiler Pluto. As Pluto is written in C, we use the C-APIs of \toolname{} for interfacing the models, to illustrating the Pipe model runners and SerDes.
We measured round-trip time using different SerDes and show the same in Fig.~\ref{fig:training-time-microbenchmarking}(e).
This integration opens new opportunities for ML-based polyhedral optimizations, including autoscheduling and tile size selection.

\section{Discussion}
\label{sec:discussion}
Let us now study the ease of integrating \toolname{} with compiler optimizations. 

\subsection{Lines of Code}
In Tab.~\ref{tab:loc-mr}, we show the number of additional Lines of Code (LOC) to integrate \toolname{} with different compiler optimizations.
We observe a significant reduction in LOC compared to the original published works.


\begin{table}
\caption{LOC to integrate model runners. \grpc shows LOC for API calls and RPC; Values in parenthesis indicate LOC in protobuf specification. Other serdes do not need any additional code.}
\label{tab:loc-mr}
\footnotesize
\begin{tabular}{lllll}
\hline
\multicolumn{1}{c}{\textbf{Optimizations}} & \multicolumn{1}{c}{\textbf{Original}} & \multicolumn{1}{c}{\textbf{\grpc}} & \multicolumn{1}{c}{\textbf{\pipe}} & \multicolumn{1}{c}{\textbf{\onnx}}       \\ \hline
\posetrl           & -    & \multicolumn{1}{r}{3+3 (4)}         & \multicolumn{1}{r}{3}         & \multicolumn{1}{r}{3} \\
\loopdist  & 65 & \multicolumn{1}{r}{3+3 (5)}         & \multicolumn{1}{r}{4}         & \multicolumn{1}{r}{3}  \\
\rlforreal & 75 & \multicolumn{1}{r}{10+3 (28)} & \multicolumn{1}{r}{4} & \multicolumn{1}{r}{3}\\
\bottomrule
\end{tabular}%
\vspace{-\baselineskip}
\end{table}
We do not compare with the size of the published version of \posetrl, as its model was not integrated with the compiler.
With Loop distribution and \rlforreal, the effort of writing Python wrappers and invoking protobuf and gRPC is completely removed.
Among the available model runners and \serdesntt, only \grpc{}, \onnx and \protobuf involve (small) additional codes to handle RPC, environment, and \protobuf messages.
It is pertinent to note that \toolname{} removes the tedious work of managing dependencies like gRPC and Python wrappers which was otherwise necessary.

\subsection{Impact on binary size, compile time and memory}
In Tab.~\ref{tab:overhead}, we show the compile time, binary size and average resident set size (RSS) used during compilation of Clang 10 with/without \toolname{}. 
The difference in binary size is $\approx 80$KB, while the average RSS value differs by $400$KB with the release build time increasing only by a few seconds.
\toolname{} incurs only a negligible overhead in terms of binary size, compile time and memory upon statically linking with the production version of Clang. 
\vspace{-1\baselineskip}
\begin{table}[h!tb]
\caption{Comparisons of time taken to build clang and final binary size with/without \toolname}
\label{tab:overhead}
\resizebox{\columnwidth}{!}{%
\begin{tabular}{lcc}
\toprule
\multicolumn{1}{c}{\textbf{Characteristics}} & \multicolumn{1}{c}{\textbf{Native Clang}} & \multicolumn{1}{c}{\textbf{Clang with \toolname}} \\ \midrule
Compilation Time                             &               5m 7s                            & 5m 15s                              \\
Binary Size                                  &               102.79 MB                       & 102.87 MB                               \\
Average RSS                                  &               1.5538 GB                       & 1.5542 GB                                \\ \bottomrule
\end{tabular}%
}
\vspace{-\baselineskip}
\end{table}

\subsection{Additional dependencies}
The current version of \toolname{} is implemented in C++17 and Python 3.10. 
While Clang17.X uses C++17, Clang10.X uses C++14. 
We updated the build system of Clang 10 to use C++17 and fixed the issues arising from the migration of earlier experiments on \posetrl, \loopdist, and \rlforreal. We were able to use Clang 17 for Inliner experiments.
Though \toolname{} itself does not introduce any dependency, model runners do: \grpcRunner, \onnxRunner, and \protoSerDes require \grpc{}, the ONNX C++ Runtime, and \protobuf setups respectively.

\subsection{Characterization}
As discussed earlier, different model runners exhibit different characteristics.
During deployment, neither of the inter-process model runners offer multi-threaded compilation upon running a single model instance.
It could be done by instantiating multiple model instances but this would consume unreasonable amounts of memory.
The in-process model runners however do not face this problem.
Though there is a separate serialization overhead involved with \grpc and pipe model runners, they are handled automatically \emph{without the involvement} of the developer.
Due to the nature of inter-process communication, there is a possibility of encountering communication errors arising from network and compiler crashes. We handle such cases as explained in Sec.~\ref{sec:errors}.
We summarize these characteristics in Tab.~\ref{tab:modelrunner-characteristics}.

\begin{table}[h!tb]
\centering
\caption{Characteristics of different model runners}
\label{tab:modelrunner-characteristics}
\resizebox{0.45\textwidth}{!}{%
\begin{tabular}{lcccc}
\toprule
\multicolumn{1}{c}{\textbf{Characteristics}} & \textbf{gRPC} & \textbf{Pipes} & \textbf{ONNX} & \textbf{TF} \\
\midrule
\begin{tabular}[c]{@{}l@{}}Multithreaded Compilation\end{tabular} & \xmark & \xmark & \cmark & \cmark \\
\begin{tabular}[c]{@{}l@{}}Distributed Training\end{tabular} & \cmark & \xmark & - & - \\ 
\begin{tabular}[c]{@{}l@{}}Need for separate model process\end{tabular} & \cmark & \cmark & \xmark & \xmark \\
Autoserialization & \cmark & \cmark & - & - \\
\begin{tabular}[c]{@{}l@{}}Communication Fidelity\end{tabular} & \xmark & \xmark & \cmark & \cmark \\
\begin{tabular}[c]{@{}l@{}}ML Framework agnostic\end{tabular} & \cmark & \cmark & \cmark & \xmark \\
\begin{tabular}[c]{@{}l@{}}Additional code by compiler writer\end{tabular} & $\mathrm{Y}$ & $\mathrm{N}$ & $\mathrm{Y}$ & $\mathrm{N}$ \\
\begin{tabular}[c]{@{}l@{}}Serialization Requirement\end{tabular} & $\mathrm{Y}$ & $\mathrm{Y}$ & - & - \\
Time overhead & $\mathrm{Y}$ & $\mathrm{Y}$ & $\mathrm{N}$ & $\mathrm{N}$ \\
\bottomrule
\end{tabular}%
}
\end{table}

\subsection{Limitations}
As mentioned earlier, not all model runners are compatible with all ML models due to the nature of the underlying libraries. For instance, Tensorflow AOT compilation supports any Tensorflow or JAX model, but not PyTorch. Also, upon exporting the inliner model from TensorFlow to ONNX, we encountered an operator (\texttt{TFL-Bucketize}\footnote{\url{https://www.tensorflow.org/mlir/tfl\_ops\#tflbucketize\_tflbucketizeop}}) that is not supported by ONNX. To handle such cases, the ONNX runtime allows registering custom operators.
Once exported, the models can be used seamlessly without restriction. 

Similarly, protobuf does not natively support C runtime. Hence, our C APIs do not support using the gRPC model runner with protobuf serialization. The current TF AOT compilation generates C++ code thereby making it not usable directly with C. This issue can be mitigated by using TF C-APIs instead of using AOT models.

\section{Related Work}
\label{sec:relatedwork}

RL environments for compilers come closest to our work, such as CompilerGym~\cite{cummins2021compilergym}, PolyGym~\cite{polygym2021pact}, Supersonic~\cite{supersonicRLgym2022cc}.
These primarily aim at facilitating research and reproducibility, which are only two of the broader ambitions of our research (e.g., deployment, programmable compiler interface, finer-grained interaction).
CompilerGym internally calls the compiler APIs from a C++ wrapper, and the communication between the Python model and the wrapper is established by predefined gRPC methods. This limits the functionality to only the APIs supported by the library and a particular compiler version with which the library is compatible. 
Supersonic~\cite{supersonicRLgym2022cc} also uses the CompilerGym way of interfacing via gRPC. 
And, to our understanding, PolyGym~\cite{polygym2021pact} does not provide a programmable compiler interface.

The gym libraries and \toolname{} solve different problems; the former facilitates research and training, while our library aims to facilitate different interfaces for communication. 
We envision \toolname{} to supplement these gym environments by providing a variety of options for more diverse, finer-grained, and framework-independent interfacing of ML models with compilers facilitating the transition from research to production.

\section{Conclusions}
\label{sec:conclusions}

We present \toolname{}, a modular and extensible library to integrate ML models within compiler optimizations. 
It provides inter/in-process model runners with different serialization options to support both training/deployment scenarios.
We show that a model and compiler pass can be integrated with only $3$ lines of code, while also enabling very deep interleaving of RL-based algorithms like \rlforreal, as well as leaner and production-friendly optimizations like function inlining.

Our library exposes C++/C and Python APIs for integration with compilers and ML frameworks respectively.
We considered multiple ML frameworks (\tf, PyTorch, RLlib), both feature-based and embedding-based representations, multiple compilers (and versions) written in different languages to show versatility and suitability of \toolname{} on research and production environments. 
We will open-source the library and artifacts with extensive documentation.


\bibliographystyle{alpha}
\bibliography{references}

\end{document}